\documentclass[runningheads,a4paper]{llncs}

\usepackage{oz}
\usepackage{color}
\usepackage{soul}
\usepackage{url}
\usepackage{graphicx}

\usepackage{citesort}

\begin{document}

\mainmatter  

\title{Towards rational and minimal change propagation in model evolution}

\titlerunning{Towards rational and minimal change propagation in model evolution}

%
%
%

\author{Hoa Khanh Dam \and Aditya Ghose}


\institute{School of Computer Science and Software Engineering\\
University of Wollongong\\
Northfields Av, Wollongong, NSW 2522, Australia\\
\email{hoa@uow.edu.au}, \email{aditya@uow.edu.au}}

%
%

\maketitle


\begin{abstract}
A critical issue in the evolution of software models is change propagation: given a primary change that is made to a model in order to meet a new or changed requirement, what additional secondary changes are needed to maintain consistency within the model, and between the model and other models in the system? In practice, there are many ways of propagating changes to fix a given inconsistency, and how to justify and automate the selection between such change options remains a critical challenge. In this paper, we propose a number of postulates, inspired by the mature belief revision theory, that a change propagation process should satisfy to be considered rational and minimal. Such postulates enable us to reason about selecting alternative change options, and consequently to develop a machinery that automatically performs this task. We further argue that a possible implementation of such a change propagation process can be considered as a classical state space search in which each state represents a snapshot of the model in the process. This view naturally reflects the cascading nature of change propagation, where each change can require further changes to be made.
\end{abstract}

\section{Introduction}



The ever-changing business environment demands constant and rapid evolution of software and consequently, change is inevitable if software systems are to remain useful. A key aspect of software maintenance and evolution is change propagation: given a set of \emph{primary changes} that have been made to software, what additional, secondary, changes are needed to maintain consistency within the system? For example, when adding a message to a UML sequence diagram, a corresponding method may need to be added to a class diagram. The \emph{secondary changes} may themselves result in new inconsistencies, which may also lead to additional changes and so on. Change propagation is very important in the process of maintaining and evolving a software system. The software engineer must ensure that the change is correctly propagated, and that the software does not contain any inconsistencies.

A large body of work in software maintenance and evolution focuses mainly on dealing with changes in source code. However, the recent emergence of model driven development has better recognised the importance of models in the software development process. Therefore, it has become increasingly important and relevant to provide support for evolution at the level of models \cite{Mens2008}, as evidenced by an emerging number of approaches (e.g. \cite{Mens00,Egyed08,Briand06} and a brief review presented in section \ref{sect-related-work}) proposed to deal with this issue. Most of the existing work addresses change propagation from the consistency maintenance perspective: change propagation can be done by detecting and fixing inconsistent relationships in a model, which are caused by primary changes made to the model.

There are many ways of propagating changes to implement a change request and maintain consistency in the model, especially in the case which the model is relatively large. Therefore, dealing with change propagation in model evolution currently faces two significant challenges: (a) how to justify whether the change request is actually implemented correctly; and (b) how to achieve effective automated change propagation since a manual version of this task would be costly, labour-intensive and in some cases error-prone, especially for complex systems. Some of the recent work have addressed these issues to a certain extent, e.g. the work of Egyed \emph{et. al.} \cite{Egyed08} or our recent proposal  \cite{HoaICSM2010} of a mechanism for fixing inconsistencies in UML models by automatically generating change options from a metamodel and consistency constraints. However, there is an emerging need for a framework that enables the reasoning of selection between alternative change options and thus the development of a machinery that automates this process.

The mature belief revision community, which spans across the areas of research in philosophy, database and artificial intelligence (AI), has faced similar issues. Belief revision (or also referred to as belief change) is the process of changing a knowledge base (i.e. beliefs) to acquire a new piece of information in such a way that the knowledge base (KB) is consistent. Extensive work in the belief revision literature has proposed techniques to automatic adaptation of a KB to new knowledge, without human intervention in the process. In this context, changes are implemented under the guiding principles that they should be both rational and minimal. Such principles are formalized as a number of postulates in the seminal AGM belief revision model\footnote{The AGM postulates are named after the names of their proponents, Alchourrón, Gärdenfors, and Makinson.} \cite{AlchourronGM85}.

We propose to tackle the aforementioned issues in model evolution based on the notions and ideas that have been developed in the belief revision literature. In the context of design models, we can view a metamodel (that a model needs to conform to) as a language and the model is a knowledge base written in that language. Therefore, belief change intuitions may serve as an inspiration (or may be directly applicable in some cases) for developing solutions to similar problems faced by current research in model evolution. In fact, using ideas from the belief revision, our earlier work \cite{Ghose1999} has defined a principled process for evolving requirements specifications and models, with minimal computation cost and user intervention. In this paper, we focus on defining an abstract change propagation framework for model evolution, similarly to the AGM framework for belief changes. This framework consists of a set of postulates that can be used to justify a selection of change options in the change propagation process. In addition, the key principle underlying this framework is the Principle of Minimal Change: when we change our model to resolve inconsistencies, we want to make a minimal change, i.e. retain as much information from the original model. This is due to the fact that developing models is generally costly and unnecessary losses of information in such models should therefore be avoided.

We then argue that a possible implementation of a change propagation process that satisfies those postulates can be developed by adapting existing search techniques in the AI literature. More specifically, we view change propagation in the evolution of a model as a state-space search in which each node represents a snapshot (i.e. state) of the model that corresponds to a step in the change propagation process. In each state, the model contains either inconsistent or consistent dependencies. A state is changed into the next one by changes to the model which may resolve some inconsistent dependencies but may also break other consistent dependencies. This view explicitly reflects the cascading nature of change propagation, where each change (primary or secondary) can require further changes to be made. In this view and under the principle of minimality, the shortest path from an initial node to the goal node is the change propagation process that satisfies the set of postulates which constraint the search.


The structure of this paper is as follows. In the next section, we give a brief overview of the state of the arts in change propagation for models. In section \ref{sect-framework}, we present an abstract change propagation framework inspired by the AGM model in belief revision theory. Section \ref{sect-search} serves to discuss the implementation of such a change propagation process using classical search techniques. We then conclude and outline some future directions of our work.

\section{Change propagation in model evolution: state of the arts}\label{sect-related-work}

In this section we attempt to briefly review the current research in supporting change propagation in model evolution. Due to space limitations, we only highlight some of the major approaches to  change propagation by detecting and resolving inconsistencies caused by changes.


The survey presented in \cite{Spanoudakis01} review various techniques and methods that have been proposed to address different activities of the consistency management process including: detecting overlaps between software models, detecting inconsistencies, identifying the source, the cause and the impact of inconsistencies, and resolving inconsistencies. The influential \emph{Viewpoints} framework \cite{Finkelstein1994} supports the use of multiple perspectives in system development by allowing explicit ``viewpoints'' containing partial specifications, which are expressed and developed using different representation styles and development strategies. In their framework, inconsistencies arising between individual viewpoints are detected by translating into a uniform logical language. Such inconsistencies are resolved by having meta-level inconsistency handling rules.

As UML has become the de facto notation for  object-oriented software development, most research work in consistency management has focused on problems relating to consistency between UML diagrams and models (e.g. a range of work presented in \cite{UML2003,UML2002}). Several approaches strive to define fully formal semantics for UML by extending its current metamodel and applying well-formedness constraints to the model (e.g. \cite{Bodeveix02}). Other approaches transform UML specifications to some mathematical formalism (e.g. Description Logic as in \cite{Mens2005book}) rely on the well-specified consistency checking mechanism of the underlying mathematical formalisms. Recently, Egyed \cite{Egyed06} proposed a very efficient approach to check inconsistencies in UML models. His approach scales up to large, industrial UML models by tracking which entities are used to check each consistency rule, and then using this information to determine which rules might be affected by a change, and only re-evaluate these rules.

Previous work on change propagation within UML models has mostly focused on fixing inconsistencies, with most work aiming to automate inconsistency resolution by having pre-defined resolution rules (e.g. \cite{Liu02}) or by identifying specific change impact rules for all types of changes (e.g. \cite{Briand06}). However, these approaches suffer from the correctness and completeness issue.  Since the rules are developed manually by the user, there is no guarantee that these rules are complete (i.e. that they generate all possible inconsistency resolutions) and correct (i.e. that the resolutions actually fix a corresponding inconsistency). In addition, a significant effort is required to manually hard-code such rules when the number of consistency constraints increases or changes.

The work in \cite{Nentwich03} addresses this issue by proposing an approach to
automatically generate repair actions for consistency constraints expressed in
xlinkit.  Recent work by Egyed  \textit{et al.} \cite{Egyed08}
proposes a mechanism for fixing inconsistencies in UML design models by
automatically generating a set of concrete changes. Their approach uses
pre-defined choice generation functions, which compute possible values for
locations in the model, for instance, possible new names for a method. The
generated options are checked against the constraints and are rejected if they
do not in fact repair the constraint, or if they cause new constraint
violations. However, this work has several major limitations. Firstly, they
consider only a single change at a time, and consequently do not take into
account the cases where a single change may not resolve all inconsistencies, or
may even temporarily introduce new ones before reaching a consistent state.
Secondly, the choice generation functions are written by hand, and may not be
complete, meaning that the approach is incomplete: it only considers a subset
of the possible ways of repairing a given constraint violation.  Finally, their
approach does not consider the creation of model elements, which in our opinion
is an important part of change propagation. Our recent work \cite{HoaICSM2010,HoaJAAMAS2011,Hoa2008PhD} overcomes the above issues by automatically generating inconsistency resolutions for UML design models.  Consistency constraints are specified using the Object Constraint Language (OCL) \cite{OCL20}, and possible inconsistency resolutions are represented in the form of abstract repair plans.

Change propagation takes place not only within a model but also between models at different levels of abstractions or expressed in different languages. In the context of model driven software development, target models are obtained from source models using model transformations. In this context, one of the issues related to model evolution is that changes in the source model should be propagated to the target model and vice versa. There are several approaches to deal with this issue. The first approach simply re-transforms the changed source model to produce a new version of the target model for each successive update made to the source model. A more efficient approach (e.g. \cite{Hearnden2006}), usually referred to as incremental transformation, ensures that subsequent changes made to the source model cause appropriate updates on the target model. Both of these approaches, however, assume that the target model was not changed when it is updated. As a result, when following these approaches any modifications that may be made to the target model will be lost the next time a model transformation takes place. Several approaches have been proposed to deal with this issue by either merging models or maintain a record of traceability links between the source and target model elements (e.g. \cite{Ivkovic04}).




Change propagation is also related to  change impact analysis, which is defined as ``\emph{identifying the potential consequences of a change, or estimating what needs to be modified to accomplish a change}'' \cite{BohnerArnold1996b}. The process of change impact analysis contains two major steps. First, the analyst examines the change request and identifies the software items (e.g. the artefacts, components, or modules) initially affected by the change. Next, the analyst identifies other items in the software that apparently have dependency relationships with the initial ones, and forms a set of impacts. Those impacted items also relate to other items and thus the impact analysis continues this process until a complete graph is obtained beginning at the selected items and ending with items on which nothing else depends. There have been a proliferation of techniques (see a recent survey in \cite{Lehnert:2011}) proposing to support change impact analysis of procedural, object-oriented systems or agent-oriented systems (seminal work presented in \cite{ArnoldBook} or more recent work such as \cite{Apiwattanapong2005,Law2003,PetrenkoR09,Maia2010,HoaICSM2011,HoaSCP2013}).
There are a few approaches that target models. The work in \cite{Kung1994} addresses how change impact analysis can be performed from a class diagram. They also discussed how object-oriented properties such as inheritance, encapsulation, polymorphism and so on affect their impact analysis. They proposed an algorithm to identify the impacted parts of the system by calculating the delta of two versions of software. Their analysis, however, only used static information, i.e. the class diagram. In addition, there has been some recent work aiming at (semi-)automated support for impact analysis of UML models. For example, the work in \cite{Knethen2003} developed a tool environment called QuaTrace that semi-automatically identifies impacts on UML models when system requirements (e.g., use case descriptions) undergo changes. Their approach is based on the establishment of traceability links between textual descriptions of use cases and UML model elements. The work of \cite{Briand06} also computes change impacts for actions for UML models but takes a different approach. They identified specific impact analysis rules (defined with the Object Constraint Language \cite{OCL20}), which are used to determine model elements that are directly or indirectly impacted by the changes. They also proposed a measure of distance between changed model elements and impacted elements in order to sort the resulting impact sets according to their probability of occurrence.

\section{An abstract framework for rational and minimal change propagation}\label{sect-framework}

Model is commonly viewed as a description of a system using a well-defined modelling language which is suitable for computer interpretation. Metamodelling is a widely used mechanism to define such a language. A metamodel is also a model which describes the abstract syntax of a modelling language, i.e. a definition of all the concepts and the relationships existing between concepts that can be used in that language. A model can have multiple views (e.g. diagams), each of which must be both syntactically and semantically consistent. On the one hand, syntactic consistency ensures that a model's view conforms to the model's abstract syntax, i.e. a metamodel, which guarantees that the overall model is well-formed. On the other hand, semantic consistency requires different views of a model to be semantically compatible (i.e. coherence). For instance, the message calling direction in a UML sequence diagram must match the class association direction in a class diagram. Such consistency requirements upon a model are often expressed using its metamodel and a set of constraints that specify conditions that a well-formed and consistent model should satisfy.

When a model is modified in order to meet a change request, typically some primary changes are made and then additional, secondary, changes are made as a result. Change propagation is the process of determining and making these secondary changes. In practice, there can be many options for resolving a given inconsistency. For instance, an inconsistency concerning a naming mismatch between a message in a sequence diagram and the operations in the message's receiver class can be resolved in different ways: either changing the message's name or changing the name of one of the operations. Choosing between those different possible repair plans can depend on various factors. For example, assume that the inconsistency is caused because the designer has renamed the message so that its name matches with a corresponding state transition. In this case, renaming an operation is possibly more preferable from the designer's perspective. The above example has shown that the cause of inconsistency can play a role in determining which repair plans should be chosen. Other elements that can influence repair plan selection include user dependent factors such as the designer's style and preferences. In addition, since we view inconsistency resolution in the context of change propagation, we have to consider the side-effects of a resolution of on other constraints. They can be negative effects, i.e. breaking a constraint, or positive effects, i.e. resolving a violated constraint. As a result, the selection of repair plans is also dependent on their side-effects.

Formulating all of those factors is not generally feasible, thus a completely automated mechanism for selecting change propagation options is not appropriate. On the other hand, it is not desirable for our change propagation mechanism to just give the designer all of the possible options. We therefore propose an abstract framework that will drive the selection of change options. We formulate a set of postulates that a ``rational'' change propagation operator needs to satisfy. The underlying motivation of these postulates (inspired by the AGM theory in belief revision) is that when we change our model, we want to make a \emph{minimal change}, i.e. retain as much information from the original model. The design process to create models is generally costly and unnecessary losses of information in such models should therefore be avoided.

In a typical change propagation process, the designer is guided by a change request which can be fulfilled by making a given change to the system's design. Based on the intuitions and ideas from the AGM model \cite{AlchourronGM85}, we propose the following postulates for a change propagation process.



\begin{enumerate}
  \item The output of a change propagation process is a well-formed and consistent model.

  \item The change propagation process must successfully implement the change request, i.e. the modified model contains the change.

  \item The change propagation process should be terminated as soon as the model becomes consistent again. We propagate changes by finding places in a model where the desired consistency constraints are violated, and fixing them until no inconsistency is left in the model.

  \item If the primary changes result in no inconsistency, then no further changes are needed.

  \item Change propagation should be analysed on the semantic level and not on the syntactical level. This means that semantically equivalent changes should lead to identical updated models. Model refactoring techniques (e.g. \cite{DBLP:journals/sosym/StraetenJM07}) mostly satisfy this condition since they change a model in such a way that its behaviour is preserved.

  \item The updated model is closest to the original model (with respect to a certain distance metric). This is also referred to as the Principle of Minimal Change, one of the most important principles in belief revision. This principle drives the selection between applicable change options: if both options resolve an inconsistency, the preferred option would change the model minimally.

\end{enumerate}

The above postulates can be used to justify an implementation of a change propagation process in the evolution of models. In the next section, we will discuss how such an implementation can be developed using classical search techniques.

\section{Change propagation as a state space search}\label{sect-search}


Similarly to Rajlich's model of change propagation of a program \cite{Rajlich97}, we view the evolution of a model as a sequence of snapshots where each snapshot represents one particular moment in the process, with some dependencies being consistent and others being inconsistent. A transition between two snapshots of the same model is triggered by a change made to an entity in the model, which may change some inconsistent dependencies into consistent ones and vice versa. This reflects the cascading nature of change propagation where performing an action to fix an inconsistency can cause further inconsistencies (side-effects) which require further actions.

From this perspective, we propose that the problem of determining a change propagation process that satisfies all the postulates set out earlier can be formulated as a state space search problem. In this state space setting, each snapshot of a model is actually represented by a state. A state in which all dependencies in the model are consistent is referred to as the ``consistent'' state, whereas a state in which some of the dependencies are not consistent is called the ``inconsistent'' state.

Furthermore, the \emph{initial state} represents the model after being modified by primary changes. In the \emph{goal state}, all dependencies in the model are consistent, i.e. the consistent state (Postulate 1). If the initial state is a consistent state, then it is also the goal state, and thus no further changes are needed to the model (Postulate 4). However, if the initial state is inconsistent, we need to find a path that takes us from the initial state to the goal state, i.e. the solution path (Postulate 3). There can be many solution paths, which reflects the fact that there are multiple ways of resolving a given inconsistency. Under the principle of minimal change (i.e. Postulate 6), we however need to find the shortest paths where the distance between nodes (i.e. states) represents how the a version of the model  (represented by a node) deviates from another version (represented by another node). It is noted that a solution path may contain intermediate, inconsistent, states, which naturally represents steps in the change propagation process.



In our previous work \cite{HoaJAAMAS2011,HoaICSM2010}, we have proposed a change propagation framework for design models based on the above approach of formulating change propagation as a state space search. This framework addresses a critical issue in state-space search: the generation of candidate states/nodes in the search is done automatically using a machinery that analyses a metamodel (specified in UML) and a set of OCL constraints. In this framework, the distance between two nodes (each of which representing the original and modified models) indicates the number of change actions that are applied to the original model to obtain the modified model. In addition, we assign each primitive action type\footnote{Changes to a model can be classified into four primitive types: creation of entities, adding and removing relationships between entities, and updating the values of attributes of entities.} with an \emph{exchange rate} (its ``basic cost''), for instance creation may have an assigned cost of 1 and deletion a cost of 3. These numbers do not correspond to any real cost, and are simply used to compare different action types. It is worth noting that the selection of the specific ``exchange rate'' values is somewhat arbitrary, and thus we allow the user to specify them.

The change propagation framework proposed in our previous however has two major issues. Firstly, in a worst case scenario our previous work performs an exhaustive search to find all the possible goal nodes. As a result, it is only applicable to small to medium models and is not scalable to larger models \cite{HoaJAAMAS2011}. A better approach would be adapting classical search techniques (e.g. best first search, A*, etc.). A challenge in this approach is defining the distance between nodes in the search space (or the cost to get from one node to another node), which correctly reflects the actual the conceptual distance between versions of a model (that the nodes represents). Another challenge is identifying admissible heuristic estimate of the distance to the goal.

Secondly, previous work (e.g. \cite{HoaAAMAS08,HoaAPSEC2010}) addresses the first challenge to some extent in terms of defining the distance between versions of a model by assigning costs to change actions. This approach is, to some extent, arbitrary, and the cheapest cost option may not always reflect the minimality of the changes. For instance, there can be many changes made to a model (resulting in a high cost), but the structure of the model may change very little or the semantics of the model may not even change (Postulate 5). Therefore, the distance between versions of a model should be defined in such a way that it captures different aspects of the model. In our recent work \cite{HoaAPSEC2010}, we have made initiatives towards this direction by proposing a set of proximity relations between versions of a service choreography model that accommodates both the structural and semantic dimensions of the model in the context of service oriented architectures. We encode a service choreography (in the form of UML activity diagrams) into semantically-annotated graphs called Semantic Process Networks (or SPNets). Structural proximity between SPNets is calculated based on the set of nodes or edges in the SPNets using either set inclusion-oriented or set cardinality-oriented measurement. On the other hand, to calculate semantic proximity, we proposed to annotate each activity in a UML activity diagram with a semantic effect and had a machinery to automatically calculate the set of effect scenarios\footnote{Due to decision and fork nodes, there may be multiple end effects of a process, each corresponding a scenario of how the process is executed.} for the process described in the activity diagram. We then used set inclusion-oriented or set cardinality-oriented measurement to define the semantic proximity between two SPNets.


\section{Conclusions and Future Work}

One of the most critical problems in software maintenance and evolution is to maintain consistency between software artefacts by propagating changes correctly (i.e. change propagation). Although many approaches have been proposed, automated change propagation is still a significant technical challenge in software engineering. We have observed many similarities between change propagation in model evolution and the belief revision in AI such as the issues of maintaining consistency, and automatically making rational and minimal changes. In this paper, we have therefore proposed an abstract framework consisting a number of postulates (inspired by the well-known AGM model in belief revision) that can be used to justify and drive a change propagation process. We have also described how an implementation of such a change propagation process can be developed using classical search techniques.


The visionary ideas that we have discussed in this paper lay out an important foundation for future work. This involves refining the proposed postulates and grounding them in a specific context (e.g. UML models). We would also need to develop a mechanism to automatically check if a change operator satisfies the postulates. Another important item for future work would involve exploring further measurement that better reflect the distance between versions of a model. Finally, a full implementation and evaluation of the framework are really necessary and are part of our future work.



%
%

\end{document}